\documentclass[12pt]{iopart}

\usepackage{iopams}
\usepackage{graphicx}
\usepackage{hyperref}
\usepackage{color}

\usepackage{bm}
\usepackage{amsfonts}
\usepackage{amsfonts}
\usepackage{latexsym}
\usepackage{subfigure}
\usepackage{lipsum}
\usepackage{cancel}

\newcommand{\hh }[1]{ \hat{\bm{#1}} }
\newcommand{\m }[1]{ \boldsymbol{#1} }

\def\eq#1{{Eq.(\ref{#1})}}

\begin{document}

\title{How many dissenters does it take to disorder  a flock?}

\author{D.~Yllanes$^{1,2}$, M. Leoni$^{1}$, M.C. Marchetti$^{1}$}

\address{$^1$ Department of Physics and Soft Matter Program, Syracuse University, Syracuse, NY, 13244}
\address{$^2$ Instituto de Biocomputaci\'on y F\'{\i}sica de Sistemas Complejos (BIFI), 50009 Zaragoza, Spain}
\date{\today}

\begin{abstract}
We consider the effect of introducing a small number of non-aligning agents
in a well-formed flock. To this end, we modify
a minimal model of active Brownian particles with purely repulsive
(excluded volume) forces to introduce an alignment interaction that will
be experienced by all the particles except for a small minority
of ``dissenters''.
We find that even a very small
fraction of dissenters  disrupts the flocking state. 
Strikingly, these motile dissenters are much 
more effective than an equal number of static obstacles
in breaking up the flock. For the studied system sizes
we obtain clear evidence of scale invariance at the 
flocking-disorder transition point and the system can be effectively 
described with a finite-size scaling formalism.
 We develop a continuum 
 model for the system
which reveals   that dissenters act like annealed noise on aligners, 
with a noise strength that grows with the persistence of the dissenters' dynamics.
\end{abstract}

\maketitle

\section{Introduction}
Flocking models inspired by the seminal work of Vicsek~\cite{Vicsek1995} have
been shown to describe organization and collective motion on many scales, from
self-motile colloids~\cite{Bricard2013} to bacteria~\cite{Copeland2009}, bird
flocks~\cite{Ballerini2008a} and human crowds~\cite{Silverberg2013,Karamouzas2014}.
In these models individual active agents are described as self-propelled
particles that tend to align their direction of motion with their neighbors, in
the presence of noise in the angular dynamics that effectively describes
``mistakes'' in the alignment. These models exhibit a non-equilibrium phase
transition from a disordered state to a flock where on average all agents are
moving in the same direction, with long-range order in the particle
velocities. The transition occurs upon decreasing the strength of the noise or
increasing the density. The order of the transition in the original Vicsek
model of point-like particles has been the subject of a long-standing debate,
but 
{it has now been established} that the transition is first order, with coexistence
and hysteresis~\cite{Chate2008,Solon2015c,Solon2015d}

Recent work has begun to consider the effect of disorder either present in the
environment in the form of physical obstacles to the
motion~\cite{Chepizhko2013,Berdahl2013,Morin2017,Pince2016,Sandor2017} or
arising from variations in the properties of individual agents or their ability
to align with
neighbors~\cite{Guttal2010,Couzin2011,Baglietto2013,Ariel2015,Copenhagen2016}. 
Both environmental disorder and disruptions
in alignment rules were found to destabilize the flocking state, in
agreement with observations in bacteria and insect swarms, where a fraction of
individuals with a decreased  production of signaling compounds or pheromones
that promote collective behavior can disrupt organization.

In this paper we consider the effect of a fraction of non-aligning agents or
``dissenters'' on a well-formed flock. Previous authors have examined the
effect of non-aligning agents on a flock 
{that is made cohesive} by attractive
interactions~\cite{Copenhagen2016}. {In this case, 
provided the cohesiveness is not too strong,
aligning agents are able to expel non-aligners and reorganize
in smaller, but still cohesive flocks. }
Our work, in contrast, focuses on the case where the
self-propelled agents only experience repulsive interactions due to volume
exclusion, in addition to alignment, {but no attractive forces}. We find that in this case even a very small
concentration of dissenters disrupts the flocking state. Additionally, this
behavior depends only weakly on the combined packing fraction of aligners and
dissenters, provided the packing fraction is large enough  that the pure
system with no dissenters is deep into the flocking state. A striking behavior
is found when comparing the effect of dissenters to that of {an equal}
concentration of static obstacles (Fig. 4). A small concentration of static
obstacles only disrupts the orientational order locally, creating small wakes
of misaligned particles downstream of the obstacles, in qualitative agreement
with experiments in colloidal rollers~\cite{Morin2017}. In contrast, the same
concentration of motile dissenters completely disrupts the flock. 
Using a hydrodynamic model of a mixture of aligners and dissenters, we show that
motile dissenters, in contrast, provide an effective annealed disorder
with finite-time correlations that can rapidly disorder the flock. This
observation could have implications for crowd control, as it suggests that
randomly distributed, but motile dissenters with persistent dynamics could be
very effective at dispersing crowds in high risk situations (see~\cite{Silverberg2013,Karamouzas2014} for studies of human ``flocks'').

In the following we begin in section~\ref{sec:model} by describing our system ---a mixture  of self-propelled
aligners and dissenters based on a minimal model of active Brownian colloids. We
then briefly summarize the physics of the pure case (aligners only) and identify
values of the parameters that result in a strongly ordered flock. This
system is then disrupted by adding a small number of dissenters, which succeed
in breaking up the alignment. The effect of these dissenters is quantified in section~\ref{sec:dissenters}
by considering high-precision simulations for several packing fractions, which leads
us to identify the fraction $p$ of dissenters required to disrupt the flock as
$p_\mathrm{c}=0.004$, independent of the total density of active particles,
 {provided again the latter is large enough to set up the flocking state.  The role of the range of the alignment interaction is discussed in \ref{sec:range}.
}
In section~\ref{sec:obstaculos} we  compare the dramatic effect of motile
dissenters to the much weaker 
disruption caused by static obstacles. 
In section~\ref{sec:FSS}
we compute the correlation length of the system and use it as the basis of
a finite-size scaling study to try to determine the order of the transition
and its critical parameters.  
Finally, in section  ~\ref{sec:continuum-main} 
{we examine the continuum equations for a mixture of aligners and dissenters (derived in \ref{sec:continuum}). We show  that dissenters act like annealed, but time-correlated, disorder and provide an analytical estimate of the shift they induce on the flocking transition. }

\section{Model and simulations}\label{sec:model}
We consider a minimal model of repulsive
active Brownian particles (ABPs)~\cite{Fily2012,Fily2014,Marchetti2016a} with 
an additional  feedback mechanism that tends to align the direction of self-propulsion
to the local velocity field~\cite{Szabo2006,Henkes2011}.
The system is composed of $N$ particles of radius $a$ in a two-dimensional
box of size $L^2$ with periodic boundary conditions. A particle $i$
is characterized by its position $\boldsymbol r_i$ and an angle $\theta_i$
that defines the direction of self-propulsion. 

The dynamics is then defined by coupled Langevin equations
\begin{eqnarray}
\boldsymbol v_i &=& \dot{\boldsymbol r}_i  = v_0 \ \hat{\boldsymbol n}_i(t) + \mu \sum_{j} \boldsymbol F_{ij}(t),\label{eq:v}\\
\dot \theta_i &=&\frac{1}{\tau} \bigl[\psi_i(t) - \theta_i(t)\bigr] + \eta_i(t),\label{eq:theta-flock}
\end{eqnarray}
The first term in the translational equation of motion represents 
the self-propulsion along a direction $\hat{\boldsymbol n}_i =(\cos
\theta_i,\sin \theta_i)$.  The second term is an excluded volume
interaction, which we model with a soft repulsive force, 
$\boldsymbol F_{ij} = \hat{\boldsymbol r}_{ij} k(2a-r_{ij})$
if $r_{ij}\leq 2a$ and $\boldsymbol F_{ij}=0$ otherwise.
The rotational equation of motion includes a  noise
term, with a random torque $\eta_i(t)$ with zero mean
and correlations $\langle \eta_i(t) \eta_j(t')\rangle = 2D_\mathrm{r}\delta_{ij}
\delta(t-t')$.  In addition to these fluctuations, the polar angle
$\theta_i$ evolves due to a  torque proportional to the angle between
$\hat{\boldsymbol n}_i$ and the instantaneous direction of motion $\psi_i$,
defined by
$\boldsymbol v_i=v_i(\cos\psi_i,\sin\psi_i)$. In other
words, as the particles collide their orientations relax towards the 
direction of the local velocity field with a lag time $\tau$ and 
a Gaussian noise of variance $ 2D_\mathrm{r}$. 

The model described by equations (\ref{eq:v}) and~(\ref{eq:theta-flock})
experiences a flocking transition if the noise is low enough (i.e., if
$\tau_\mathrm{r}=D_\mathrm{r}^{-1}$ is large compared to $\tau$) or
if the density is increased at fixed noise~\cite{Szabo2006,Henkes2011}.

Notice that alignment can be modeled in several ways {--- see,
e.g.~\cite{Weber2013} for an alternative approach with active polar hard
disks---} but the results presented herein are not very sensitive to the
details (see \ref{sec:range}  for an example with a Vicsek-type
alignment~\cite{Vicsek1995}). Likewise, whether the repulsive interaction is
introduced with a spring force or with a harder potential (such as WCA) should
not have a noticeable effect. 

In our simulations we take the radius $a=1$ of the disks as our unit
of length and set $\mu=k=1$, taking the interaction time 
$\tau_k=(\mu k)^{-1}$ as our temporal unit. The alignment lag
is also set to $\tau=1$. The self-propulsion speed is $v_0=0.01$, which is small
compared to $\mu k$ to prevent 
 particle overlap. The noise is set to $D_\mathrm{r}=0.0005$:
this results in a persistence length of $\ell_\mathrm{p}=v_0/D_\mathrm{r}=20$
and generates a strong alignment.
We will change the packing fraction $\phi$ in order to transition 
from the low-density disordered state to the high-density 
flocking state. { We consider system sizes of up to
$L=400$, which, for a typical packing fraction of $\phi=0.40$, results
in about $20000$ disks.}
\begin{figure}[tb]
\centering
\includegraphics[width=.6\linewidth]{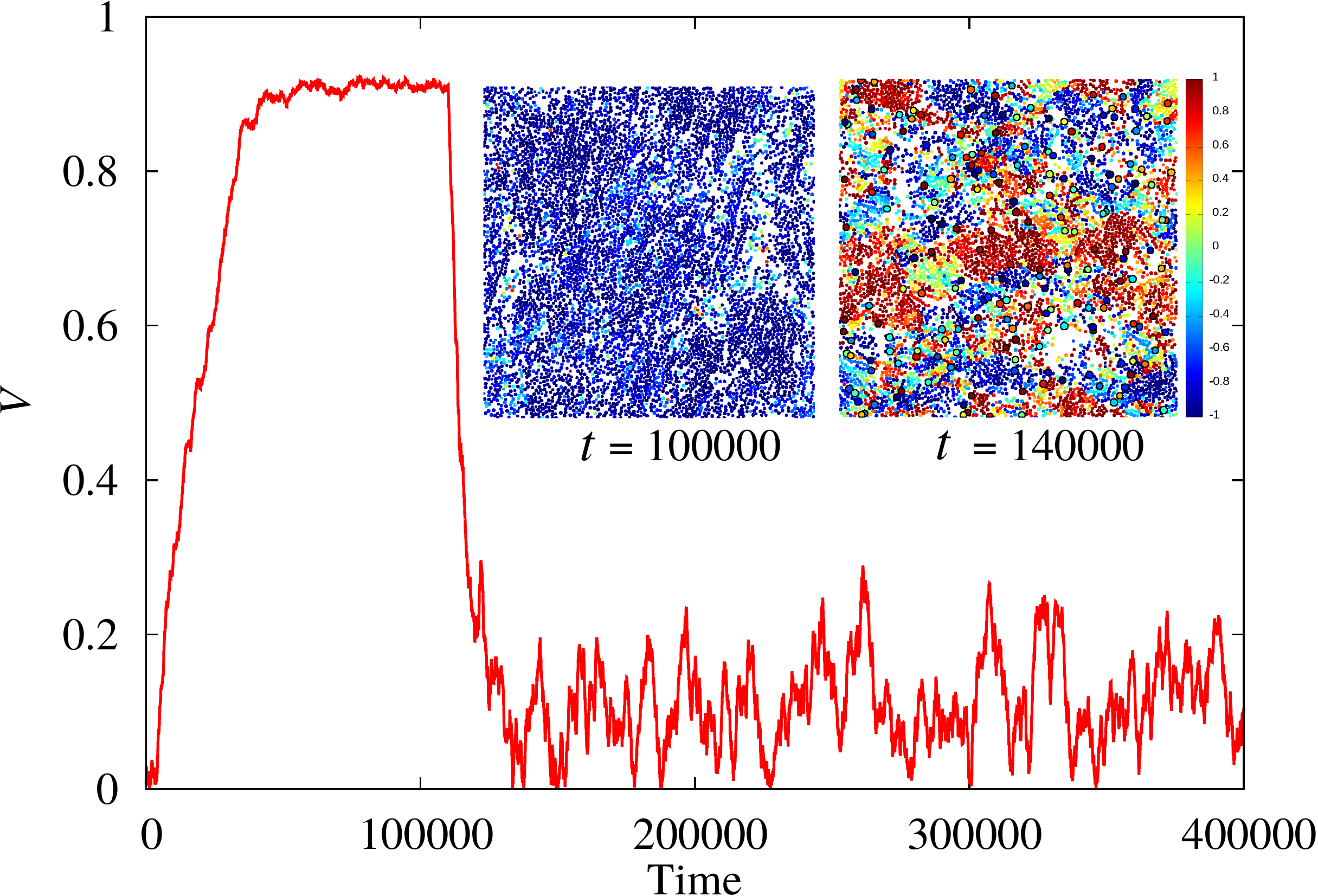}
\caption{Effect of introducing dissenters in a flocking system. The figure shows 
the time evolution of the order parameter $V$, Eq.~(\ref{eq:V}). The system is prepared 
without any dissenters and is let to form a flock. At time $t=1.1\times10^5$, we remove the alignment 
interaction from $3\%$ of the particles (i.e, turn them into dissenters). The flock is dispersed very quickly.
Two snapshots show the state of the system just before and just after introducing the dissenters. Each particle
is plotted with a color representing the cosine of its angle of motion.
 \label{fig:quench}}
\end{figure}

Now we introduce a second species in the system. Out of 
our $N$ disks, $(1-p)N$ will still be \emph{aligners}, described by Eqs.~(\ref{eq:v})
and~(\ref{eq:theta-flock}). The other $pN$ disks will be \emph{dissenters}:
they have the same characteristics as the aligners, except
for the alignment interaction. In other words, their equations
of motion are just those of standard ABPs:
\begin{eqnarray}
\boldsymbol v_i^{(\mathrm{d})}(t) &=& \dot{\boldsymbol r}^{(\mathrm{d})}_i = v_0 \ \hat{\boldsymbol n}_i(t) + \mu \sum_{j} \boldsymbol F_{ij}(t),\label{eq:v-diss}\\
\dot \theta_i^{(\mathrm{d})}(t) &=& \eta_i(t).\label{eq:theta-diss}
\end{eqnarray}
In the following we consider simulations of our combined system
of aligners and dissenters for different values of $p$ 
and $\phi$. 

\section{Effect of the dissenters}\label{sec:dissenters}
 Even a very small fraction $p$
of dissenters can have a dramatic effect on the system. This is demonstrated
in Figure~\ref{fig:quench}, where we follow the time evolution of the system
before and after introducing dissenters. In particular, we consider
the average velocity of the system as the flocking order parameter,
\begin{equation}\label{eq:V}
\boldsymbol V = \frac1N \sum_i \frac{\boldsymbol v_i}{|\boldsymbol v_i|}\,\qquad   V = | \boldsymbol V|\, .
\end{equation}
Clearly $V=1$ if all the disks are moving in exactly the same direction
and  $V=0$ if their orientations are random.

We {first consider the pure system with} $p=0$, where all the particles 
experience the alignment interaction. In the initial configuration 
the positions and orientations of the particles are random but, 
as time goes on, a stable flock develops, as evident from the growth and
saturation of $V$ around $V=0.91$ shown in Fig.~\ref{fig:quench}. At time $t=1.1\times10^5$, we turn $3\%$ of the particles
into dissenters, i.e., we switch off their alignment interaction. The effect
on the system is very strong and fast: the flock is destroyed in a very short
time (shorter than what it took to form originally). 

\begin{figure}[tb]
\centering
\includegraphics[height=.6\linewidth,angle=270]{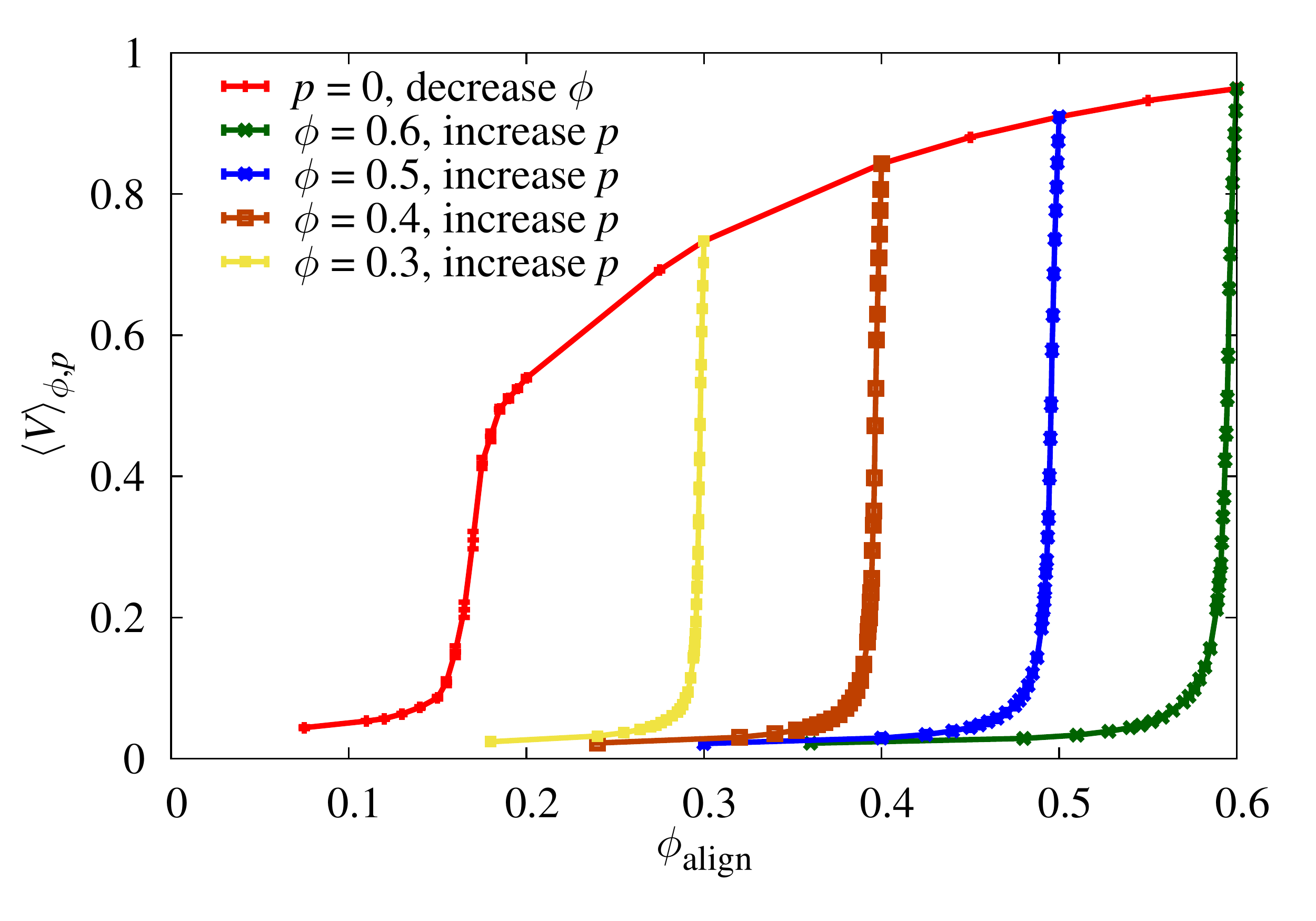}
\caption{$\langle V\rangle_{\phi,p}$ against 
the  packing fraction of 
aligners ignoring the dissenters,
$\phi_\mathrm{align} = \phi(1-p)$. The red curve shows
the flocking transition in the pure case (no dissenters, $p=0$), as
the system is diluted by changing $\phi$.
In the other curves we 
consider a fixed total packing fraction $\phi$ and 
slowly increase the fraction $p$ of dissenters until the alignment is destroyed.
\label{fig:phi-flock}}
\end{figure}

Figure~\ref{fig:phi-flock} gives a more general picture
by considering the steady-state value of $V$ for many values
of $p$ and $\phi$.  For each pair $(\phi,p)$ we follow the system up to
a time $t=5\times10^5=250D_\mathrm{r}^{-1}$. We denote by
$\langle O\rangle_{\phi,p}$ the ensemble average in the steady state
of an observable $O$, which we estimate numerically by averaging
over the last half of our simulation (the time needed to reach the steady
state is orders of magnitude shorter). 
The errors are estimated with 
a jackknife procedure (see, e.g.,~\cite{Amit2005})
 from the 
fluctuations over 100 independent runs for each set of parameters.
This method allows us to 
 compute errors in non-linear 
 functions of averaged quantities
  such as the susceptibility. 
Unless we say otherwise, all of the results in this paper are 
for a system size $L=200$.

In order to plot all the data in the same graph and also to prove
that the effect of the dissenters is much stronger than that {of simply}
diluting the system, we define as $\phi_\mathrm{align}=(1-p)\phi$, the packing
fraction of aligners ignoring the dissenters. The red
curve in Figure~\ref{fig:phi-flock} {refers to} the pure system, where we keep
$p=0$ and decrease the total packing fraction ($\phi=\phi_\mathrm{align})$.
As we can see, with our parameters we need a 
rather strong dilution in order to break our flock and cross over to the 
disordered state. In contrast, in each of the other curves we fix 
the total packing fraction $\phi$ and change $\phi_\mathrm{align}$ 
by slowly increasing the fraction $p$ of dissenters. In agreement with
Figure~\ref{fig:quench}, we see that a very small value of $p$
is enough to destroy the flock. The behavior does not seem
to depend much on the value of $\phi$.
\begin{figure}[tb]
\centering
\includegraphics[height=.6\linewidth,angle=270]{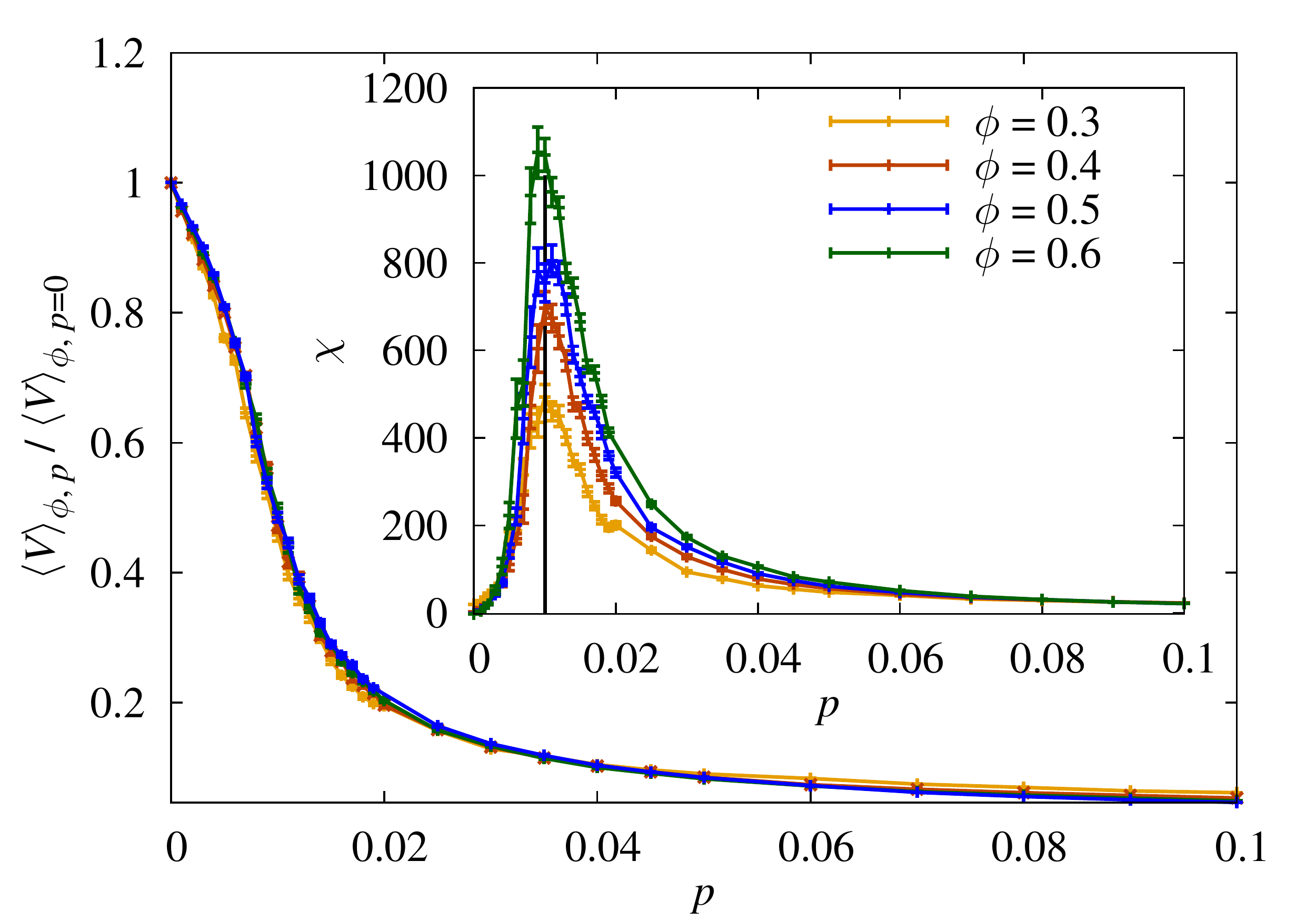}
\caption{For the data with variable $p$ in Figure~\ref{fig:phi-flock},
we plot $\langle V\rangle_p$, normalized by the value for $p=0$ for 
each packing fraction. The curves for different $\phi$ collapse, showing
that the effect of the dissenters does not depend on the density 
of the system. \emph{Inset:} We plot the susceptibility~(\ref{eq:susc})
for the data in the main panel, whose peak at 
$p_\mathrm{max}\approx 0.01$ marks the crossover between
the flocking and disordered phases.
\label{fig:V-p}}
\end{figure}
\begin{figure}[t]
\centering
\includegraphics[width=.85\linewidth]{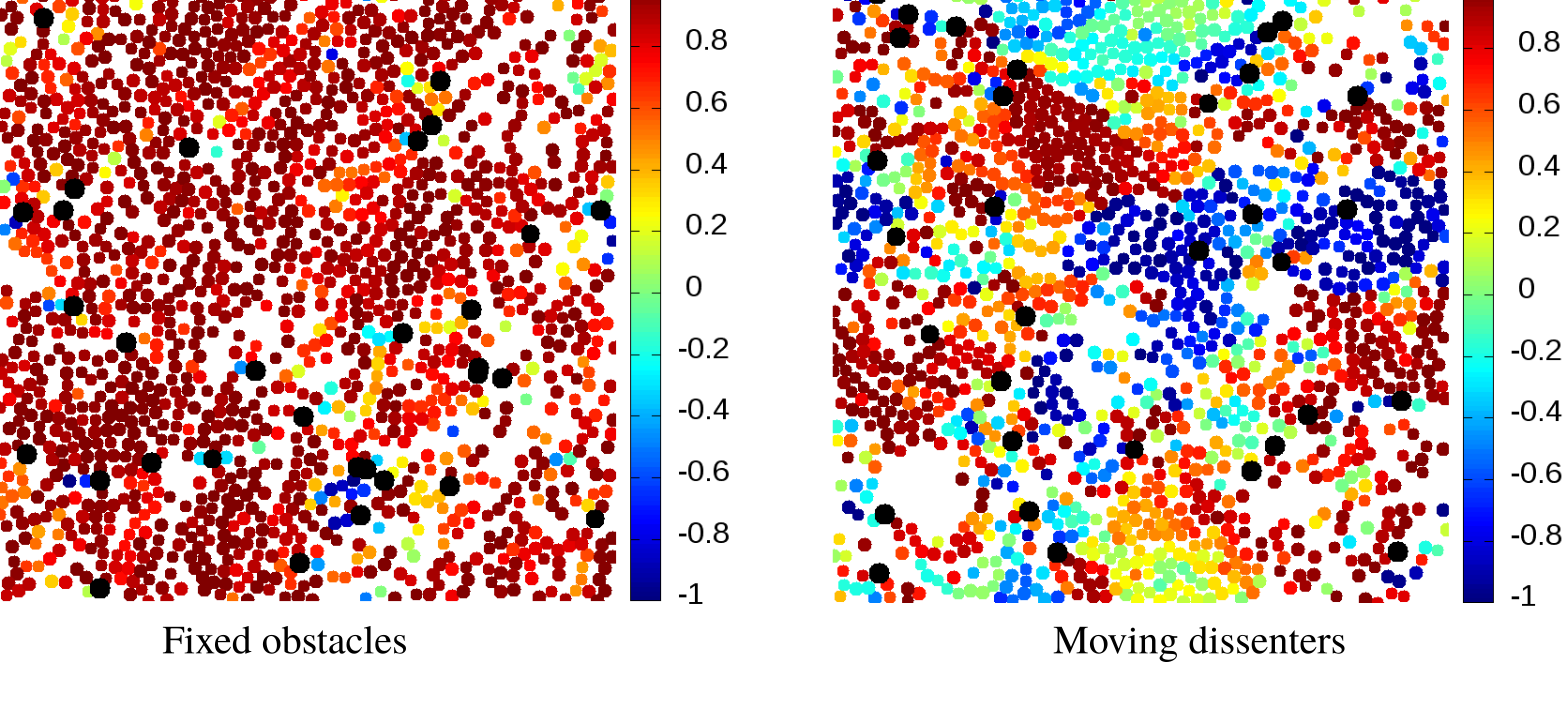}
\caption{Snapshots of 
two systems  with $\phi=0.50$ in the steady state. In the left panel we have a fraction
$p=0.03$ of static obstacles, while in the right panel we consider the same number
of moving dissenters. Both images are for  $L=100$. 
{Aligners are color coded  according to the value of 
 $\cos(\phi_i)$, as indicated in the color bar. Obstacles or dissenters are shown in black and with
size slightly larger than their actual size}. 
\label{fig:obstacles}}
\end{figure}

{To} make this statement more quantitative 
{we plot in Figure~\ref{fig:V-p} the order parameter normalized to its
 value at $p=0$, $\langle V\rangle_{\phi,p}
/\langle V\rangle_{\phi,p=0}$,} 
as a function of $p$.
All the curves collapse on top of one another, showing that (i)
a very small fraction of dissenters is enough to completely disrupt
the alignment and that (ii) this fraction does not depend on the density of
the system.
{To locate} the transition point, we consider the 
fluctuations of the order parameter,
\begin{equation}\label{eq:susc}
\chi = L^2 \bigl( \langle V^2\rangle_{\phi,p} - \langle V\rangle^2_{\phi,p}\bigr).
\end{equation}
We shall refer to $\chi$ as the susceptibility of the system,
in analogy with equilibrium systems where Eq.~(\ref{eq:susc})
is an expression of the fluctuation-dissipation theorem~\cite{Amit2005}. 
As we can see in the inset of Figure~\ref{fig:V-p}, $\chi$ has 
a maximum for $p_\mathrm{max}\approx0.01$, which signals the 
finite-size crossover between the ordered and disordered phases. We will make this 
statement more precise below, where we outline a finite-size scaling
study.

{It is interesting to compare this critical fraction 
of dissenters of $p\approx 0.01$ to the corresponding value
for the finite cohesive flocks studied in~\cite{Copenhagen2016}. In the 
latter, considering the limit of large inter-agent cohesiveness, typical
values of the critical fraction of dissenters are around $p\sim0.5$ (see
Figure~1 in~\cite{Copenhagen2016}). This fraction can even be further
increased by lowering the cohesiveness, which allows aligners to expel
dissenters and reorganise into several flocking clusters.}

\section{Static obstacles}\label{sec:obstaculos}
 The effect of passive obstacles 
in a flocking system has been considered before using different
models~\cite{Chepizhko2013,Sandor2017}. In this section we 
show that our active dissenters are much more efficient
at disrupting the alignment than static obstacles. This contrast
is illustrated  in Figure~\ref{fig:obstacles}, which shows snapshots
of our model system with dissenters ($\phi=0.50$, $p=0.03$)
in the left panel and  a system where the dissenters have been
replaced by the same concentration of static obstacles (right panel). These obstacles are just immobile
disks of the same radius as our active particles. {The collisions of the aligners 
with the static obstacles is controlled by the same repulsive force that controls their interaction with motile dissenters.} 
As we can see, for this fraction of static obstacles the  system
still maintains a high degree of alignment, while the system
with dissenters is completely disordered.

\begin{figure}[t]
\centering
\includegraphics[height=.6\linewidth,angle=270]{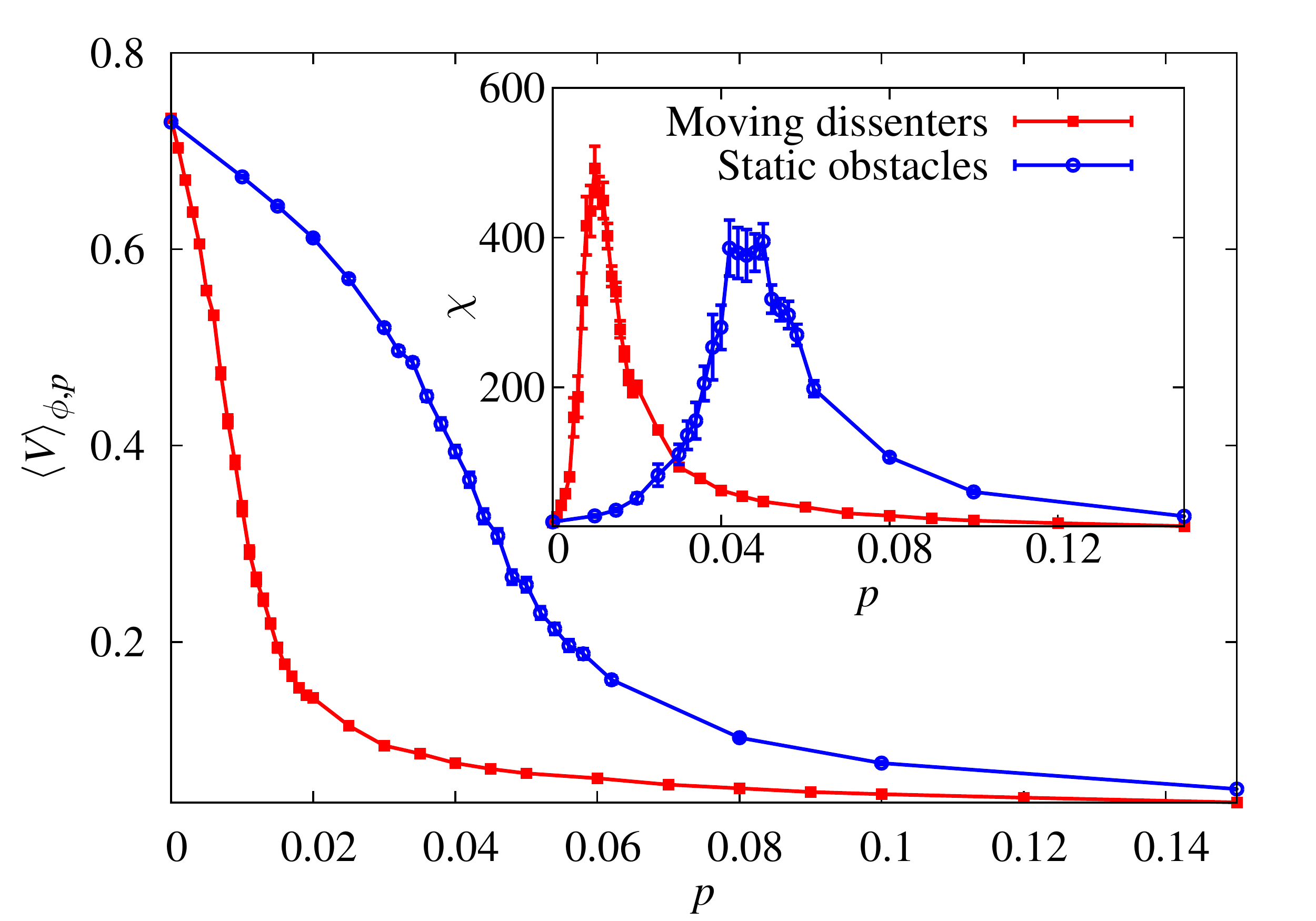}
\caption{As in Figure~\ref{fig:V-p}, but now we compare,
for $\phi=0.30$, the effect of introducing dissenters (red curve)
with that of introducing static obstacles (blue curve). The obstacles
are much less efficient at breaking up the flock, as evinced by the very 
noticeable shift in the peak of the susceptibility and in the slower decay 
of $\langle V\rangle_p$.
 \label{fig:V-obstacles}}
\end{figure}

The difference between passive obstacles and dissenters
{quantified} in Fig.~\ref{fig:V-obstacles}
 {that compares the order parameter and susceptibility for systems with static obstacles (blue) and moving dissenters (red).}
For the model with static obstacles,
the peak is at $p_\mathrm{max}^\mathrm{static}\approx 0.05$, in contrast 
with $p_\mathrm{max}\approx 0.01$ for the dissenters. In other words,
one needs approximately five times as many static obstacles as dissenters
to have an equally disruptive effect.

The contrast between static obstacles and motile dissenters
 is probably due to the  latter effectively
providing an annealed disorder with finite-time correlations, in contrast to
the weak quenched disorder of static obstacles. 
The persistent movement
of the dissenters effectively gives them a greater cross section. 
 We have tried to support this
intuition with a continuum model  presented in the following, see Section~\ref{sec:continuum-main}.

\section{The correlation length and finite-size scaling}\label{sec:FSS}
 Thus far
we have presented essentially an exploratory study of a minimal model
of flocking particles with dissenters. We have seen that a very small
number of these non-aligning particles (about $1\%$) is enough to
disrupt the flock. But this effect was observed for a single
system size ($L=200$). In order for this result to be considered
a proper (non-equilibrium) phase transition, we would need to
show that  the effect
of the dissenters is stable as we change the system size.
\begin{figure}[t]
\centering
\includegraphics[height=.6\linewidth,angle=270]{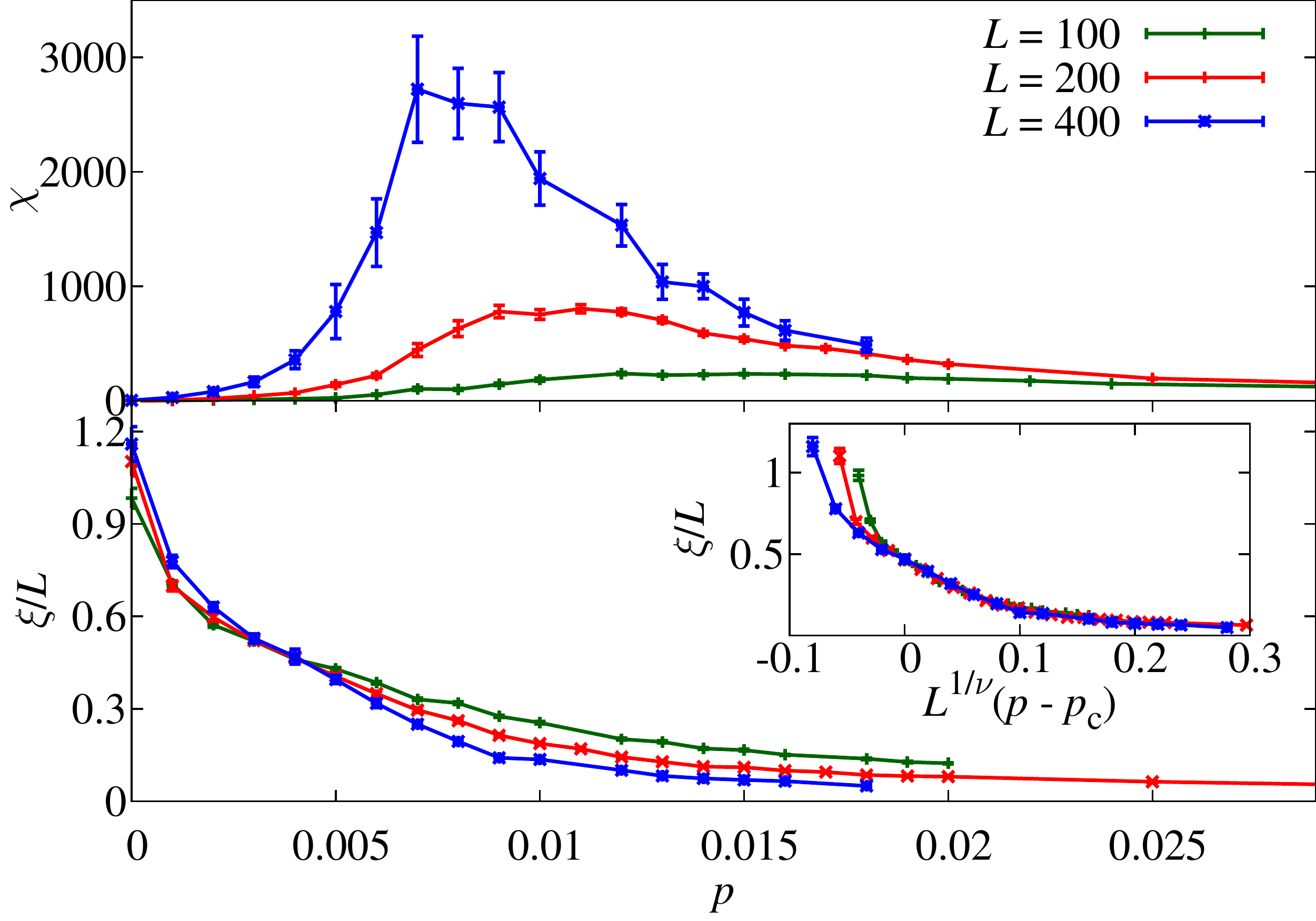}
\caption{Scaling in the system with dissenters for $\phi=0.50$.
\emph{Top:} We plot the susceptibility $\chi$, Eq.~(\ref{eq:susc}),
for $L=100,200,400$. As we increase $L$, the position of the peak shifts slightly 
to a lower $p$ and its height grows. \emph{Bottom:} Correlation 
length of the particle orientation in units of the system size. The
curves for different values of $L$ intersect at $p\approx 0.004$, marking
a second-order phase transition. \emph{Inset:} Scaling 
plot of the correlation length using $\nu=2$ and $p_\mathrm{c}=0.004$.
 \label{fig:scaling}}
\end{figure}

To this end, we have carried out additional simulations with $L=100,400$ 
for the system with dissenters and $\phi=0.50$. As we mentioned before,
the peak in the susceptibility signals the finite-size crossover between
the disordered and the flocking phases. When the system size is increased,
this crossover region becomes narrower and narrower, as the crossover
turns into a phase transition in the thermodynamic limit. To leading order,
the position of the peak should evolve as
\begin{equation}\label{eq:scaling-ji}
p_\mathrm{c} \simeq p_\mathrm{max}(L) + A L^{-1/\nu},
\end{equation}
where $A$ is a constant, $\nu$ is the correlation-length critical exponent and $p_\mathrm{c}$ 
is the transition point. This 
behavior is qualitatively reproduced in our Figure~\ref{fig:scaling}--Top.
In principle, we could fit the data in this plot to extract the critical parameters
$\nu$ and $p_\mathrm{c}$. Unfortunately, with only three system sizes such a simultaneous
fit for two parameters is not viable.

A better way to analyze a phase transition is to use the system's 
correlation length $\xi$. We begin by considering the spatial autocorrelation
of the particle orientation:
\begin{equation}\label{eq:corr}
C(\boldsymbol r) = \langle \hat {\boldsymbol n}(\boldsymbol x) \cdot \hat {\boldsymbol n}(\boldsymbol x + \boldsymbol r)\rangle_{\phi,p}\ .
\end{equation}
In order to evaluate this quantity, we first discretize the system in a lattice
with cells of size $2\times2$. The orientation $\hat{\boldsymbol n}(\boldsymbol x)$
of cell $\boldsymbol x$ is just the average of all the $\hat{\boldsymbol n}_i$ 
of particles in that cell. Then we evaluate
\newcommand{\ee}{\mathrm{e}}
\newcommand{\ii}{\mathrm{i}}
\newcommand{\dd}{\mathrm{d}}
\begin{equation}\label{eq:F} 
F(\boldsymbol k) = \biggl\langle\biggl|\sum_{\boldsymbol x} \ee^{\ii  \boldsymbol k\cdot \boldsymbol x} \hat{\boldsymbol n}(\boldsymbol x)\biggr|^2\biggr\rangle\,
\end{equation}
which is just the Fourier transform of $C(\boldsymbol x)$. From $F$ 
we can compute the 
second-moment correlation length~\cite{Cooper1982,Amit2005},
\begin{equation}\label{eq:xi}
\xi = \frac{1}{2 \sin{k_\mathrm{min}}} \left(\frac{F(0)}{F(\boldsymbol k_\mathrm{min})} -1\right)^{1/2}\, ,
\end{equation} 
where $\boldsymbol k_\mathrm{min}=(2\pi/L,0)$ is the smallest non-zero wavevector. 

In a second-order phase transition the system is scale invariant, so  the
correlation length behaves as \begin{equation}\label{eq:scaling-xi} \xi \simeq
L g\bigl( L^{1/\nu} (p-p_\mathrm{c})\bigr), \end{equation} In other words, if
we plot $\xi/L$ for our different  system sizes, the curves will intersect at
the transition point. We have done this in the bottom panel of
Figure~\ref{fig:scaling}, which shows that the system is indeed scale
invariant, with a critical point of $p_\mathrm{c}\approx 0.004$. In addition,
we can find $\nu$ by looking for the value that produces the best collapse
in~(\ref{eq:scaling-xi}). With our data, this is obtained for $\nu\approx2$
(although we cannot obtain a very precise determination). This scaling plot is
shown in the inset to Figure~\ref{fig:scaling}. Notice that the points for
$p=0$ are already out of the finite-size scaling (FSS) region and do not
collapse but this is expected, because these points are deep into the ordered
phase, where $F(0)$ diverges.  These values of $\nu$ and $p_\mathrm{c}$ are
consistent with our data for $p_\mathrm{max}(L)$ and Eq.~(\ref{eq:scaling-ji}).

{ Our simulations show  evidence of a continuous flocking
transition. On the other hand, it has been established recently that the
flocking transition in the pure Vicsek model is first order, with coexistence
and hysteresis~\cite{Chate2008,Solon2015c,Solon2015d}. At the level of a
continuum description the first order nature of the Vicsek flocking transition
arises from the density dependence of the term linear in polarization in the
polarization equation. In contrast, in models where the alignment is with
topological neighbors rather than nearest neighbors, this term does not depend
on density and the transition is continuous~\cite{Chen2012,Peshkov2012}. In our
model the alignment with the particle's own velocity is density dependent and
we would therefore expect the transition to be first order.  On the other hand,
establishing the first-order nature of the flocking transition of point
particles in the Vicsek model has required simulations with very large numbers
of particles~\cite{Solon2015c}, with a  crossover size that depends on the
details of the model and parameters~\cite{Chate2008}. Below that crossover
there is a wide range
of system sizes where the transition looks continuous and finite-size scaling 
holds~\cite{Attanasi2014,Baglietto2008}.  When steric repulsion is included in
the model, it becomes even harder to see clustering and band formation ---the
hallmark of the first order flocking transition (see \ref{sec:excluded})--- as
the fact that particles cannot overlap forces them to distribute more uniformly
throughout the system.  In fact, to the best of our knowledge, all studies of
the flocking transition that find a first-order behavior have been carried out
with point particles.  Indeed,
we have found that in the presence of both dissenters and static obstacles
density fluctuations are much weaker in our model that includes steric
repulsion that in a model of point particles  of the type studied in
Ref.~\cite{Baglietto2013}. We show examples of this different behavior in
\ref{sec:excluded}. 

It is therefore likely that the scaling behavior observed in our work is a
finite-size effect, and that for large system sizes the transition is first
order. Unfortunately, with numerical simulations alone it is impossible to
differentiate between an asymptotic regime and a pre-asymptotic one that would
have a crossover at very large system sizes, well beyond those relevant to
practical applications such as human crowds.  }

\section{Continuum model}
\label{sec:continuum-main}

To gain insight on the  picture emerging from our  simulations, we have developed
  a continuum model  that describes the system on length scales large compared to the particle size and 
   time scales long  compared to those controlling the microscopic dynamics.
In this limit we describe the mixture of flocking (aligning) and dissenter agents in terms of the local number density of aligners
 and dissenters at position $\m x$ and time $t$, $\rho(\m x, t)$ and  $\rho^{D}(\m x, t)$, respectively, and the corresponding 
 polarization densities    $\m P(\m x, t)$ and $\m P^D(\m x, t)$. 
In this continuum model the net polarization $\m P(\m x, t)+\m P^D(\m x, t)$  serves as the order parameter for the flocking transition. 

\subsection{Hydrodynamics of a mixture of aligners and dissenters}
   The continuum equations have been derived via a standard coarse-graining procedure for a simplified 
   {continuous-time} Vicsek model where all agents are treated as point particles that align their polarization to that of of their neighbors (see, e.g.,~\cite{Farrell2012}). 
   The derivation is outlined in~\ref{sec:continuum}. 
   {
   We stress that the model used for the derivation of {the} hydrodynamic theory
    differs from the one used in the simulations as it 
 considers point particles
   that align with the mean polarization of their neighbors, not with their own velocity.
   }
   While the form of the hydrodynamic equations does not depend on the specific
form of the microscopic dynamics, the latter does of course affect the
expression of the parameters in the equations. The use of a 
   {continuous-time Vicsek model} greatly simplifies the algebra. In
addition, {the assumption of point particles} allows a direct comparison of the
effect of dissenters with that of static obstacles as 
   of course 
   point static obstacle would have no effect on the organization of the aligners.
The continuum equations obtained in \ref{sec:continuum} are 
 given by 
\begin{eqnarray}
\fl  \partial_t \rho  = 
-  v_0 \boldsymbol{\nabla} \cdot  \m P \;, & 
\nonumber\\
 \fl  \partial_t 
 \m P   +\lambda   \left(\m P \cdot\bm\nabla\right) \m P 
  = 
 \left[\alpha(\rho)-\beta|\m P+\m P^D|^2\right]\m P+ \gamma\rho\m P^D
 -  \frac{v_0}{2} \boldsymbol{\nabla}  \rho 
 +     K_A  \nabla^2  \m P  
 +\sqrt{{2\Lambda}\rho} ~\m f\;, 
 &
 \nonumber
\\
\fl  \partial_t \rho_D   = 
  - v_D \boldsymbol{\nabla} \cdot  \m P^D\;, &
 \nonumber  \\
\fl  \partial_t  \m P^D 
  = 
  -D_\mathrm{r}   \m P^D - \frac{v_D}{2} \boldsymbol{\nabla}  \rho^D + K_D  \nabla^2  \m P^D+ \sqrt{{2\Lambda}\rho^D}~\m f \;,
&
\label{eq:dot-pa}
\end{eqnarray}
where for generality we have distinguished the self-propulsion speed  $v_D$  of
 dissenters from that of aligners given by $v_0$. 
 Here we have added a 
 white noise  term 
  $\bm f$   with zero mean
  and correlations $\langle f_i(\m x, t) f_j(\m x', t')
  \rangle =   \delta_{ij} \delta(t-t')\delta(\m x- \m x')$, needed    to compute correlation functions, and we estimate $\Lambda\sim D_r$. 
 {The polarization decay rate }$\alpha(\rho) $ changes sign at a critical density $\rho_c$ and 
$\gamma, \beta > 0$.
 For the microscopic model described in~\ref{sec:micro-model}, 
  the various parameters in Eqs.~(\ref{eq:dot-pa}) are expressed 
 in terms of the  {rotational diffusion rate $D_r$ and the rate $J$ at which particles align with their neighbors' polarization (see Eqs.~ (\ref{eq-app-micro})).}
 {Although the alignment interaction used in the derivation of hydrodynamics differs form that employed in the numerical simulations described in Eqs.~\ref{eq:theta-flock}, both models exhibit a flocking transition driven by alignment at low noise and high density, and we can estimate $J\sim 1/\tau$. This is also supported by the discussion of the role of the range of the alignment interaction presented in 
\ref{sec:range} 
 }
For the continuous-time Vicsek model used in \ref{sec:continuum} we find $\alpha(\rho)=Ja^2\rho-D_r$, $\beta=J^2a^4/2 D_r$ and $\gamma=Ja^2$, 
{with $a$ the range of the aligning interaction.} 
The continuum equations then yield a transition at $\rho_c=D_r/(Ja^2)$ from an isotropic state with vanishing mean polarization at low density to a polarized or flocking state with $\rho=\rho_0$, $\rho_D=\rho^D_{0}$,  $\m P^D_{0}=0$ and $\m P_0=\mathbf{\hat{x}}P_0$, and $P_0=\sqrt{\alpha(\rho_0)/\beta}$. The flocking state breaks rotational symmetry
 spontaneously. Without loss of generality we have then chosen the $x$ axis along the flocking direction. The dissenters never order and their presence does not 
 affect the mean-field transition. 
Finally, the stiffnesses  
$K_D$ and  $K_A$ are controlled by the interplay of self-propulsion and rotational noise, with  $K_A =  \frac{v^2_0}{16 D_\mathrm{r}}  $ and $K_D =   \frac{v_D^2}{16 D_\mathrm{r}} $,
and the advective parameter is 
$\lambda = \frac{3 v_0 J a^2 }{8  D_\mathrm{r}} $. 
We have neglected other advective nonlinearities that do not affect the behavior deep in the ordered phase.

\subsection{Correlation functions}
~\label{sec:correl-function}
We now examine the effect of dissenters on the correlation function of fluctuations of the order
 parameter away from the direction of order.  {We linearize the equations}
deep in the ordered phase {by letting} $\delta\rho=\rho-\rho_0$ and
$\delta \m P=\hh x \delta P_x  + \hh y P_0 \delta \theta$.
For simplicity in the following we set $P_0= \rho_0$. Using the linearized equations given in \ref{sec:corr-continuum} and eliminating $\delta P_x$ in favor of density fluctuations,  
we evaluate the correlation function of the Fourier components of the angular fluctuations $\theta(\m q,\omega)=\int_{\m r, t} e^{-i\omega t+i\m q\cdot\m x}~\delta\theta (\m x,t)$, with the result
\begin{equation}
\label{eq:theta-corr}
\fl \langle |{\theta}(\m q, \omega)|^2 \rangle = 
 \frac{  ( \omega + v_1  q_\parallel )^2\left[  2  \Lambda/\rho_0 + \gamma^2  ~\langle| P^D_{y}(\m q,\omega)|^2\rangle\right]}
 { ( \omega + v_1  q_\parallel )^2   K^2_A   q^4 
+   \left[( \omega + v_1  q_\parallel ) ( \omega   + \rho_0 \lambda q_\parallel )  -    v^2_0  q^2_\perp/2 \right]^2 },
\end{equation}
where
\begin{equation}
\label{eq:pd-corr}
\langle| P^D_{y}(\m q, \omega)|^2 \rangle \simeq
 \frac{  2\rho^D_{0}\Lambda }
 {\omega^2+D_r^2} \;,
\end{equation}
{
and
$v_1 =  v_0 [  \alpha'(\rho_0) P_0] /[2 \alpha(\rho_0) ] \approx v_0 $, 
where the prime $'$ denotes a derivative with respect to density and
the approximate equality holds deep into the flocking state. 
It is evident from Eqs.~(\ref{eq:theta-corr}) and (\ref{eq:pd-corr}) that the dissenters play the role of noise that is correlated over the time scale $\tau_r = D_r^{-1}$.
{We examine} the long wavelength behavior of the equal time correlation function, given by 
$\langle|\theta(\m q)|^2\rangle=\int_\omega \langle |{\theta}(\m q, \omega)|^2 \rangle$, where $\int_\omega ...=\int \frac{d\omega}{2\pi}...$ {for wavevectors along the direction of broken symmetry, 
i.e., by letting 
$q_\perp=0$. In this limit density and angle fluctuations decouple. Incorporating a finite
$q_\perp \neq 0$ changes the angular angular dependence of the correlation function, but not the leading long wavelength behavior  \cite{Toner2012}.
Furthermore, a finite $q_\perp$
 affect the contributions from} annealed noise and dissenters in the same way.
}
The details of the calculation are given in \ref{sec:corr-continuum}, with the result
 \begin{equation}
\langle |  {\theta}(\m q) |^2  \rangle \simeq
\frac{ 
 \Lambda }{ \rho_0  K_A  q^2_\parallel}
+\frac{ 
 \rho^D_{0}\Lambda 
\gamma^2
  }{ 
  K_A  q^2_\parallel D^2_r  
} 
\simeq
 \frac{  \Lambda}{\rho_0K_A   q^2_\parallel}  
 \left[1+\rho_0\rho_0^Da^4\left( \frac{J}{D_r} \right)^2\right]\;.
 \label{eq:correlation-result}
\end{equation}
The first term on the right hand side of Eq.   (\ref{eq:correlation-result}) is the result for the pure system, while the second is the contribution from the dissenters.
Both terms have the same behavior at large length scales, with the dissenters enhancing the noise strength by an amount proportional to $(\gamma/D_r)^2\sim(J/D_r)^2$.
Although the correlation function here diverges as $1/q^2$ at small wavevectors, as expected for the fluctuations associated with the Goldstone modes of the broken symmetry phase in two dimensions, it is known that nonlinearities stabilize the  polar flocks  \cite{Toner2012}. 
Our numerics suggest that a small fraction of dissenters enhances the effective noise, hence shifting the order-disorder transition. This is also supported by the mean field calculation presented in the next section.

In contrast, in the limit of point particles considered here, static obstacles would have simply no effect as they would not couple at all to our active agents, 
leaving the flocking state unperturbed. In a system of finite size particles with steric interactions, an areal density $\rho^D_{0}$ of static obstacles
described by  
{
quenched disorder 
with correlations 
 $ \langle  F_i(\m x, t)   F_j(\m x', t') \rangle \sim \beta^2_0 \nabla_i \nabla_j \delta(\m x -  \m x') $,
yields
}
  angular  spatial fluctuations \cite{Morin2017} 
 \begin{equation}
 \langle | {\theta}(\m q, \omega) |^2 \rangle\sim\frac{  \beta^2_0 
 k^2_\perp  }{
  \rho^2_0 \lambda^2 k^2_\parallel 
}
\label{eq:delta-theta3}
\end{equation}
that, although anisotropic, remain finite at large scale. Self-propelled agents essentially only interact with static obstacles for a time inversely proportional to their self-propulsion speed. In contrast, our dissenters travel at the same speed as the aligners 
 and their  influence persists over times of order  
 $\tau_r$, during which aligners can align with dissenters provided  $\tau_r  >  J^{-1} $.

\subsection{Shift of the order-disorder transition}

The effect of dissenters on the flocking transition can be quantified by a simple mean-field argument. 
To do this, we consider the homogeneous equation for the polarization of the aligners, and replace the terms 
coupling to the polarization of dissenters by their mean-field value. Denoting by $\bar{\m P} $
 the homogeneous aligners polarization, we obtain 
\begin{equation}
\partial_t  \bar{\m P} =
 \left[\alpha(\rho_0)
-\beta  \langle |\m P^D |^2 \rangle\right]
\bar{\m P}
  -\beta \bar{\m P}^2 \bar{\m P}\;,
\label{eq:result1}
\end{equation}
where we recall $\alpha(\rho_0)=Ja^2-D_r$. 
Since $ \langle |\m P^D |^2 \rangle > 0$,   
 dissenters  suppress the transition by shifting $\alpha(\rho_0)$ to smaller values, or, equivalently, 
  enhancing the alignment rate $J$ and suppressing
  the 
  noise $D_r$.  
  This effect can be
  quantified by estimating  $ \langle |\m P^D |^2 \rangle$ 
  by using \eq{eq:P-Fourier} as 
  \begin{equation}
\langle |\m P^D(\m r, t)|^2  \rangle =  
\int  \frac{ \dd^2 \m q}{(2\pi)^2}
 \int  \frac{ \dd \omega }{2\pi}
\int  
\left \langle
|\m P^D (\m q, \omega) |^2
\right\rangle .
\label{eq:P-Fourier2}
\end{equation}
The integral  over  $\m q$    has to be regularized by introducing a short wavelength cutoff. 
To 
estimate this integral we examine the  limit of small  dissenter speed   $v_D \sim 0$ 
and 
\begin{equation}
\left \langle
|\m P^D (\m q, \omega) |^2
\right\rangle  =  \frac{4 \Lambda \rho^D_0}{\omega^2 + D^2_r }.
\end{equation}
Using a short wavelength cutoff of the order of the average separation among aligners in \eq{eq:P-Fourier2},
we obtain 
\begin{equation}
\langle |\m P^D(\m r, t)|^2  \rangle
=
\frac{\rho_0  \rho^D_0}{
2  \pi 
}\,.
\end{equation}
The correction can then be  
   recast as an effective rotational diffusion
   constant $D^{'}_r=D_r+\beta  \langle |\m P^D |^2 \rangle$, given by
 \begin{equation}\label{eq:Drprima}
D_\mathrm{r}' \simeq D_\mathrm{r} \left[1 +\frac{ a^4 \rho_0  \rho^D_0 }{4\pi}~\left(\frac{J}{D_{r}}\right)^2 \right] \;,
\end{equation}
where
  we have used $\beta =(Ja^2)^2/2 D_r$.
If the persistence time $\tau_r$ of the dissenters is large compared to the time scale $\tau\sim J^{-1}$ required for alignment, dissenters strongly enhance the effective rotational noise, driving the flocking transition to higher density. If, in contrast, $\tau_r\ll \tau$, 
($D_r \gg J$),
the dissenters have little disrupting effect on a well aligned flock as
{
 moving aligners  do not have time 
  to align with  dissenters that are rapidly changing their orientation. 
 }
As expected, the enhancement of noise also
 increases with the packing fraction of dissenters $\phi^{D}$.
For the values of $D_\mathrm{r}$ used in our simulations, the enhancement of the noise due to the dissenters 
can be very strong, even for a very low $\phi^{D}$, in agreement with our observations.
This simple estimate offers a qualitative, but not quantitative agreement with our numerics.
Finally we note that  unlike the numerical model, \eq{eq:Drprima} does not depend on the ratio $\rho^D/\rho$ but only on $\rho^D$.
This is likely due to the absence of excluded volume interactions in our analytical description. 
Note that dissenters enhance rotational noise even when $v_D = 0$. In this limit they simply provide a noisy alignment interaction.

\section{Discussion.}
 We have shown that a small number
of dissenters can break up a well-formed flock. The critical fraction
of dissenters does not seem to depend on the total density of the system
and is much lower than the number of static obstacles required 
for an equally disruptive effect. 
Such results are qualitatively understood by using a continuum
model 
for a mixture of aligners and dissenters.
For the simulated system sizes, we find evidence 
of scale invariance. Indeed, the presence
of excluded-volume interactions 
suppresses density fluctuations as compared to 
models of point particles, as shown in Appendix B. It is therefore likely that our system sizes are below the crossover ones required to observe 
band formation, and that the transition in this system is indeed first order. Establishing the nature of the flocking transition is not, however, the focus of our work. 
Our main new results are the demonstration that (i) very few dissenters, far
fewer than static obstacles, are needed to disorder a flock, and (ii) that the
fraction of dissenters that causes the flock to break is independent of the
system's total density. These results are obtained for moderate system sizes 
relevant to experimental realizations such as human crowds.

It is interesting to contrast our results with those of~\cite{Couzin2005},
where it was found that the proportion of leaders needed to guide a group to
the desired destination decreases with increasing group size. In contrast, we
find that the proportion of dissenters needed to break a flock does not depend
on the flock size. There is, however, an important difference between the
leaders modeled in Ref.~\cite{Couzin2005} and our dissenters, in that 
leaders, like dissenters, are not influenced by the rest of the pack, but,
unlike dissenters, maintain a fixed, as opposed to random, orientation. 

{
Previous work with static obstacles has found that tuning particle
properties such as their repulsion~\cite{Quint2015} or their
noise~\cite{Chepizhko2013} can have a non-monotonic effect
on their order and that, therefore, there are optimal
values that maximize flocking in a disordered environment.
These results, are, however, not directly applicable 
to the case with moving dissenters. For instance, we have found
that changing the intensity of rotational noise (but using
the same value for aligners and dissenters) has almost no effect
on the critical fraction of dissenters, as long as the noise
is low enough to permit flocking in the $p=0$ limit. On the other
hand, the fact that the dissenters rapidly diffuse across the system, which is
therefore effectively homogeneous averaged over intermediate time scales, 
probably means that there is no optimal aligner noise 
for a fixed value of the dissenter noise (except in the limit
of slow-moving dissenters). Therefore, trying to find 
a simple mechanism that would mitigate the 
effect of dissenters remains an interesting open question.}

We believe that our system based on active Brownian particles 
with excluded volume interactions is especially well suited to model
collective phenomena in densely packed human crowds (in contrast 
to the more common models with point-like particles). Therefore, our
results could have implications for crowd control in high-risk situations, 
suggesting that a small number of randomly placed motile 
agents could be very effective at, for instance, dispersing human 
avalanches.

\section*{Acknowledgments} 
M.C.M. was supported by NSF-DMR-1305184 and NSFDMR-1609208 
and by the NSF IGERT program through award NSF-DGE-1068780.
 M.L. acknowledges financial support from the ICAM Branch contributions.
D.Y. acknowledges support by MINECO/FEDER (Spain)
through Grant No. FIS2015-65078-C2-1-P.  We acknowledge 
the resources and assistance provided by BIFI-ZCAM (Universidad de Zaragoza),
where we carried out our simulations on the Memento and Cierzo
supercomputers. All authors thank the Soft Matter Program
at Syracuse University for additional support.

\appendix
\section{Increasing the alignment range}\label{sec:range}
\begin{figure}[t]
\centering
\includegraphics[height=.6\linewidth,angle=270]{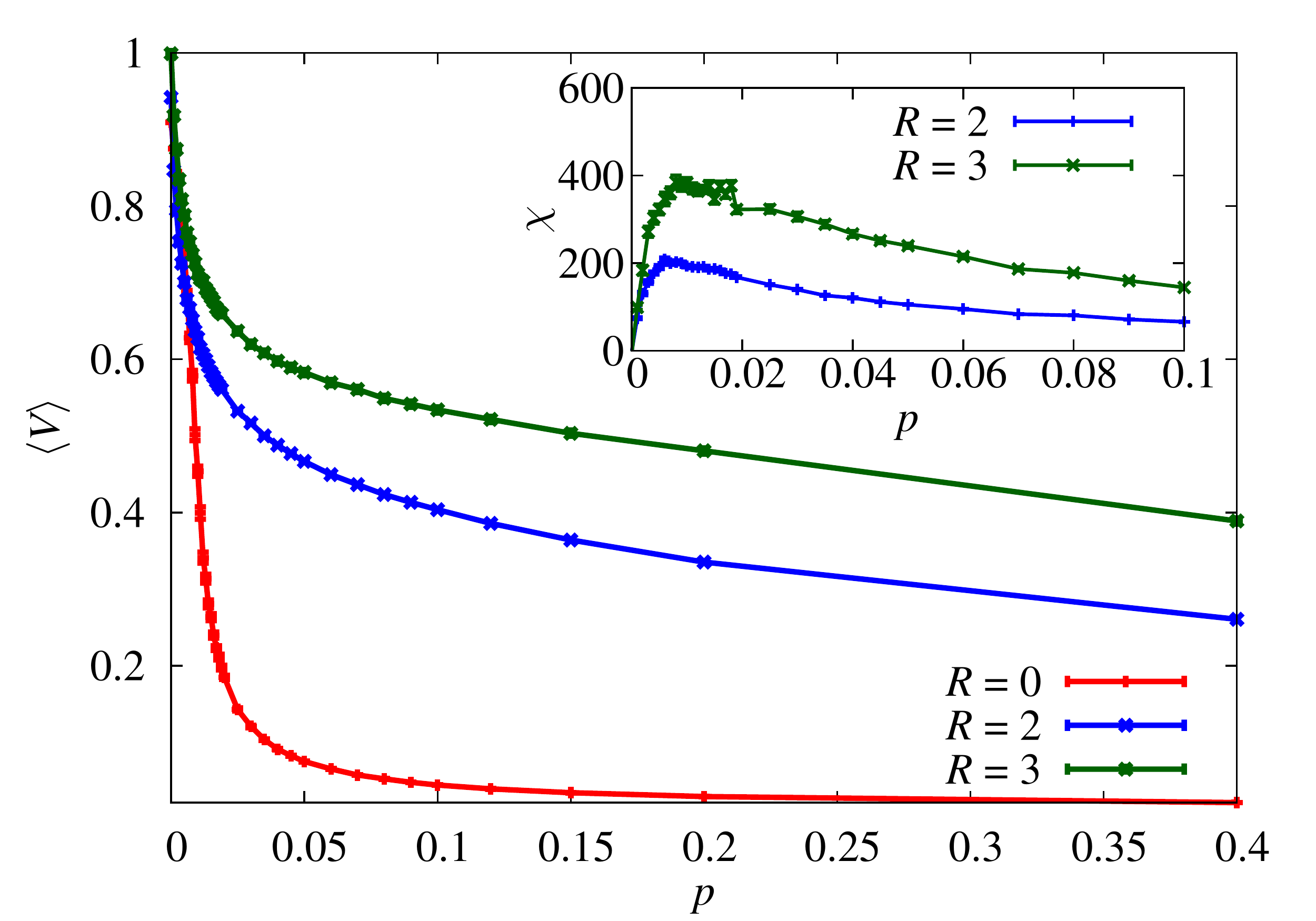}
\caption{The figure displays the effect of increasing the alignment 
range using the modified model of Eq.~(\ref{eq:theta-range}).
All the data are for $\phi=0.50$. The curve for $R=0$ corresponds
to the model previously shown in Figure~\ref{fig:V-p}. Increasing 
the alignment range makes it more difficult for the dissenters to completely 
break the flock, resulting in a slower decay of $\langle V\rangle_{\phi,p}$.
The effect of just a few dissenters is, however, still very strong and 
the peak in the susceptibility (i.e., the transition point) hardly moves
with respect to the $R=0$ case (see \emph{Inset}).
\label{fig:V-range}}
\end{figure}

{ In this paper, we have considered a model where particles align
their self-propulsion speed with their own velocity, which is in turn
determined by interactions with other particles.}
This seems the most natural
choice for our model of finite-size disks. 
Many flocking studies use, however, a different alignment interaction, 
with each particle trying to
relax to the average orientation of its neighbors within a finite range $R$
(this is the case in the Vicsek model~\cite{Vicsek1995},
which consists of point particles).
In this Appendix we consider this alternative and stronger alignment 
mechanism by modifying our equations of motion 
to read
\begin{eqnarray}\label{eq:theta-range}
\dot \theta_i  &=& \frac{1}{\tau} (\psi^R_i - \theta_i) + \eta_i,\\
\psi^R_i &=& \frac1{\sum_{j: r_{ij}\leq R} 1}{\sum_{j: r_{ij}\leq R} \arctan \frac{\sin \psi_j}{\cos \psi_j} }.
\end{eqnarray}
Notice that we are using the average angle of motion
$\psi^R_i$ instead of the average orientation $\theta^R_i$.
We do this so the limit $R=0$ coincides with our 
original model. Choosing $\theta^R_i$ or $\psi^R_i$
makes no practical difference, since at a given time
most particles are not interacting with any other
and therefore have $\theta_i=\psi_i$.

The result of using this alternative alignment mechanism is shown 
in Figure~\ref{fig:V-range}. Clearly, for our finite system of $L=200$,
this enhanced alignment makes it more difficult for the dissenters 
to completely destroy the alignment. However, the derivative
of $\langle V\rangle_{\phi,p}$ at the origin is still very 
large. More importantly, the peak of the susceptibility 
for this model is still at $p_\mathrm{max}\approx0.01$
for $R=2,3$, as in the $R=0$ case. Therefore, the introduction 
of this stronger alignment has no noticeable effect on 
the crossover between the ordered and disordered phases. It 
merely results in a wider $\chi$ peak, with longer tails and, therefore,
stronger finite-size effects (which makes sense, considering that
the system size in units of the alignment range is smaller).

{
\section{The effect of excluded-volume interactions}\label{sec:excluded}
Most numerical studies of flocking, starting with the seminal paper by Vicsek
et al.~\cite{Vicsek1995}, consider point particles. We have, instead, opted for
a model which, in addition to alignment, has excluded-volume interactions,
because we believe that this in an important factor for practical applications
(such as human crowds), even though it makes simulations harder. In this 
appendix we briefly explain the qualitative difference between these two kinds 
of flocking systems. 

In order to simulate flocking point particles, we drop the excluded-volume
interaction from the modified model presented in \ref{sec:range} (we need the model with explicit alignment range $R$ since there will be no collisions). That is, aligners
will move according to the equations 
\begin{eqnarray}\label{eq:theta-point}
\boldsymbol v_i &=& \dot{\boldsymbol r}_i  = v_0 \ \hat{\boldsymbol n}_i(t) ,\\
\dot \theta_i  &=& \frac{1}{\tau} (\psi^R_i - \theta_i) + \eta_i,\\
\psi^R_i &=& \frac1{\sum_{j: r_{ij}\leq R} 1}{\sum_{j: r_{ij}\leq R} \arctan \frac{\sin \psi_j}{\cos \psi_j} },
\end{eqnarray}
This is very similar to the model considered in~\cite{Baglietto2013}.

Let us first consider the case $p=0$, for which we plot two snapshot of the
system in panels (a) and (b)  of Figure~\ref{fig:excluded_volume}.  Panel (a)
considers the case with particle radius $a=1$, just like in the rest of the
paper. Since there are no dissenters the particles are moving in basically the
same direction, with some small deviations due to occasional random collisions.
Crucially, the disks are distributed essentially homogeneously throughout
the simulation box. In panel (b), on the other hand, we show an equal number of
flocking point particles. The configuration is now very different: since there
is nothing to keep the particles apart, the flock is much more concentrated and
the local density is very heterogeneous. In fact, the flock is starting to form
a well-defined band, as has been widely reported for the Vicsek model (see,
e.g.,~\cite{Solon2015c}).

Once we introduce dissenters, the differences between finite disks and point particles
become even more striking. On panel (c) of Figure~\ref{fig:excluded_volume} we
show a snapshot of the system with excluded volume and a high concentration of
dissenters ($p=0.1$). Now there is no flock, but the excluded-volume
interactions still force the particles to space themselves uniformly. 
Panel (d) shows the corresponding configuration for the same number of point particles
and $p=0.1$. Now the dissenters have forced the aligners to aggregate in tiny but
very concentrated clusters, each moving in a random direction, while the dissenters 
themselves are naturally still distributed uniformly throughout the system.

Notice that this comparison explains why, unlike in ~\cite{Baglietto2013} or recent
studies of the flocking transition in the Vicsek model, we find scale 
invariance at the transition, since the steric repulsion prevents concentrated bands
of particles from forming and keeps density fluctuations small.

\begin{figure}
\centering
\includegraphics[height=.9\linewidth,angle=270]{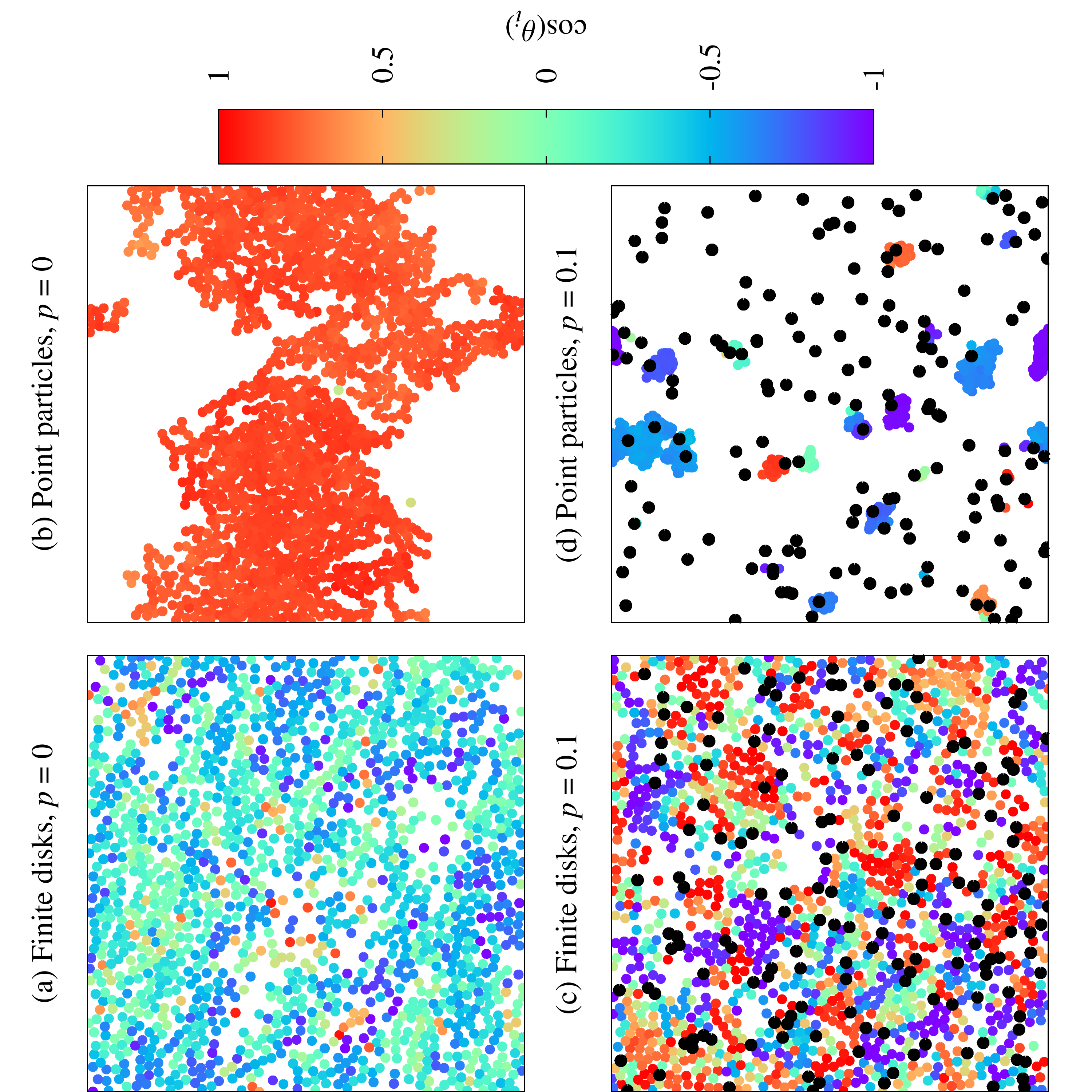}
\caption{ Comparison of the system's behavior with and without 
excluded volume interactions. We show snapshots of steady-state 
configurations for four cases. The left column of plots, panels (a) and (c),
consider  self-propelled disks of radius $a=1$, just as in the rest
of the paper. Each disk is plotted with a color given by its 
instantaneous orientation (the packing fraction is $\phi=0.5$). In panel (a), there are no dissenters ($p=0$)
and most particles have the same orientation. In panel (c), on the  
other hand, we have introduced a fraction $p=0.1$ of dissenters (in black)
and the system is disordered, as shown by the very different disk colors.
The right column considers analogous cases, but now for an equal number
of point particles (which are nevertheless plotted as disks for comparison
purposes). Panel (b) shows the pure system, with a very ordered flock
that, unlike in panel (a), is forming a band. In panel (d), the dissenters
have caused the aligners to aggregate in several small and very concentrated
clusters. In all cases we have used the modified model of \ref{sec:range} with $R=2$.
\label{fig:excluded_volume}}
\end{figure}

}

{
\section{Hydrodynamics of a mixture of aligners and dissenters}
\label{sec:continuum}

Here we 
derive the hydrodynamic equations for aligners and  dissenters
using 
{a continuous-time Vicsek model} of point particles.
The model contains two simplifications as compared to the one used in simulations:
(i) we neglect excluded volume
interactions among the particles;
 and
(ii) 
  aligners  align with  the orientation
 of neighboring particles,
  instead of aligning with 
  their own direction of motion. 
  These  simplifications allow  us to  carry out  the derivation of the continuum equations
analytically.
    Note that 
 the findings of  \ref{sec:range} }indicate
    that  the details of the alignment affect the dynamics   only 
    quantitatively, but not qualitatively.

\subsection{Derivation of the continuum equations}
\label{sec:micro-model}

Aligners are located at positions $\m r_i$ for $i = 1,
\ldots N$ and 
{are self-propelled at speed} $v_0$ in the direction $\hh n_i = ( \cos
\theta_i, \sin \theta_i)$.   Likewise, dissenters  
{located at positions  $\m r^D_i$,  for $i = 1, \ldots M$, have self-propulsion speed} 
$v_D$ along  $\hh n^D_i = ( \cos \theta^D_i, \sin \theta^D_i)$.  
 Aligners align with each
other and also with dissenters,  whereas dissenters cannot  align.  
{The dynamics  of the system is governed by}
\begin{eqnarray}
\dot{\m r}_i &=&  v_0 \hh n_i\;,
\qquad \dot{\theta}_i = \eta_i +  \sum_{j \neq i} J_{ij} \sin(\theta_j -\theta_i ) +   \sum^M_{j=1} J_{ij}  \sin(\theta^D_j- \theta_i)\;, \nonumber \\
\dot{\m r}^D_i  &=&  v_D \hh n^D_i\;, \qquad \dot{\theta}^D_i = \eta_i\;.
\label{eq-app-micro}
\end{eqnarray}
The alignment couplings have the form $  J_{ij}   =  2 Ja^2  \delta(\m r_i -\m
r_j)$, which describe  contact interactions between particles of size $a$.  
This determines a range of interaction comparable to that of our simulations.
As in the numerical model discussed in section~\ref{sec:model}, $\eta_i$ are stochastic terms describing white noise with correlations
$ \langle \eta_i(t) \eta_j(t') \rangle = 2 D_\mathrm{r} \delta_{ij} \delta(t-t') $.

We introduce the one-particle density of aligners (dissenters) describing the probability of 
finding an aligner (dissenter) at  position $\m r$ ($\m r^D$) moving in the direction $\hh n =( \cos \theta ,^D  \sin \theta ) $  ($\hh n^D =( \cos \theta^D ,  \sin \theta^D ) $) at time $t$ as
\begin{eqnarray}
& c(\m r, \theta, t) =  \left\langle \frac{1}{N} \sum^N_{i=1} \delta(\m r- \m r_i(t)) \delta(\theta -\theta_i(t)) \right\rangle, \nonumber \\
& c^D(\m r^D, \theta^D, t) = \left\langle \frac{1}{M} \sum^M_{i=1} \delta(\m r^D- \m r^D_i(t)) \delta(\theta^D -\theta^D_i(t)) \right\rangle .
\end{eqnarray}
The  continuum equations for  aligners and dissenters can be derived by coarse-graining the
 microscopic equations (\ref{eq-app-micro}), following a standard procedure (see, e.g.,~\cite{Farrell2012}).
First, one obtains  {noise-averaged} Smoluchowski equations for both aligners and dissenters of the form 
\begin{eqnarray}
\fl \left(\partial_t +v_0 \hh n \cdot  \bm{\nabla} \right)c(\m r, \theta, t) =&
D_\mathrm{r} \partial_\theta^2  c(\m r, \theta, t)
- 2  J a^2 \partial_\theta    \int \dd\theta'  \sin(\theta'- \theta)  c(\m r, \theta, t) c(\m r, \theta', t) \;,
 \label{eq:dot-cA} \nonumber \\
&
-    2 J a^2\partial_\theta    \int \dd\theta^D  \sin(\theta^D-\theta) c(\m r, \theta, t) c^D(\m r^D, \theta^D, t) ], 
\nonumber \\
&  \left(\partial_t +v_D \hh n^D \cdot  \bm{\nabla}\right)c^D(\m r^D, \theta^D, t) =  D_\mathrm{r} \partial^2_{\theta^D} ~ c^D (\m r^D, \theta^D, t)\;.
\nonumber 
\end{eqnarray}
{To obtain} equations for density and polarization
 {we now consider the angular moment of the probability densities, given by}
\begin{equation}
f^\alpha_n (\m x, t)= \int \dd\theta \left. \right. \ee^{n \ii \theta } c^\alpha(\m x, \theta, t)\;,
\label{eq:moments}
\end{equation}
with $\alpha = A, D$ labeling aligners or dissenters.
  We  indicate  complex conjugated
 using an overbar.
     The  first few moments 
     are related to { density and polarization density}, 
with
\begin{eqnarray}
&f^A_0 = \rho\;,~~~~ f^A_1 =  P_x + i P_y\;,\nonumber\\
&f^D_0 = \rho^D\;,~~~~f^D_1 = P^D_x + i P^D_y \;.
\end{eqnarray}
{denoting for simplicity the zeroth moments by $\rho$ and $\rho^D$,} the  equations for the first  few moments 
{of the aligners density}
are given by 
\begin{eqnarray}
& \fl  \partial_t \rho + 
 \frac{ v_0 }{2}    \left[ \partial_z f^A_1 + \partial_{\overline{z} }  \overline{f^A_1 } \right] =0 ,
\label{eq:dot-moments} \\
&\fl \partial_t f^A_1 
+ \frac{v_0}{2}   \left[  \partial_z  f^A_2 +  \partial_{\overline{z} } \rho  \right] 
=-D_r f^A_1  +   J a^2 [ 
\{   
 - \rho  f^A_1 
+   
 \overline{ f^{A}_1 } f^{A}_2
\}  
  +
 \{    
- \rho   f^D_1
+  \overline{f^{D}_1}   f^{A}_2
\} 
],
\nonumber \\
&\fl \partial_t f^A_2+ \frac{v_0}{2} {\left[\partial_{z }f^A_3 +\partial_{\overline{z}}   \overline{f^A_1}\right]}
=-4 D_r f^A_2
- 2 J a^2  
\{   
   f^A_1    f^A_1
+  f^A_1   f^D_1
\}, 
\nonumber 
\end{eqnarray}
where we  have defined 
$\partial_{{z} }= 
 \frac{\partial}{ \partial_x}  -i  \frac{\partial}{ \partial_y}  $
and
 $
\partial_{\overline{z} }= 
 \frac{\partial}{ \partial_x}  + i  \frac{\partial}{ \partial_y}.
$
Similarly, for the dissenters we obtain
\begin{eqnarray}
& \partial_t \rho^D +
  \frac{ v_D }{2} \left[ \partial_z f^D_1  + \partial_{\overline{z}} \overline{f^D_1 }  \right] =0\;,
\nonumber \\
& \partial_t f^D_1 
+\frac{v_D}{2}     \left[  \partial_z f^D_2 +  \partial_{\overline{z}} \rho^D   \right]
=- D_r f^D_1 
\nonumber \\
& \partial_t f^D_2 + \frac{v_D}{2} {\left[\partial_{z }f^D_3 +\partial_{\overline{z}}   \overline{f^D_1}\right]}=
-4 D_r f^D_2 .
\label{eq:dot-moments-diss}
\end{eqnarray}

 As {discussed in \cite{Bertin09}, a consistent approximation for a system with polar symmetry is obtained by neglecting all moments of order equal to or higher than $n=3$ and noting that the second moment $f^\alpha_2$ is proportional to the component of a nematic order parameter that in a system with polar interactions decays on microscopic time scales even in the ordered flocking state. We therefore neglect $\partial_t f^\alpha_2$ } in  \eq{eq:dot-moments} and \eq{eq:dot-moments-diss}
and  eliminate $f^\alpha_2$ from the dynamics 
using
\begin{eqnarray}
& f^A_2
\simeq
\frac{1}{4 D_r} \left[ - \frac{v_0}{2}  \partial_{\overline{z} }    f^A_1 - 2 J  a^2 \left(f^A_1  f^A_1 +  f^A_1    f^D_1 \right)\right]\;,
\nonumber \\
& f^D_2
\simeq
-\frac{v_D}{8 D_r}  \partial_{\overline{z} }    f^D_1\;.
\end{eqnarray}
 Replacing these expressions in 
 \eq{eq:dot-moments} and
  \eq{eq:dot-moments-diss}
 we obtain a closed system of equations for
 $f^\alpha_0$ and $f^\alpha_1$ which result in the set of hydrodynamic equations \eq{eq:dot-pa} of the main text.

\subsection{Correlation functions}
\label{sec:corr-continuum}

{To evaluate the correlation function of polarization fluctuations, we linearize the hydrodynamic} equations deep in the flocking state
by letting {$\rho=\rho_0+\delta\rho$, $\rho^D=\rho_0^D+\delta\rho^D$, $\m P^D=\delta \m P^D$ and
$\m P =  \m P_0 + \delta \m P $, with $\m P_0=P_0\hh x$}. We set
$P_0= \rho_0$.
and write $\delta \m P =  [\hh x \delta P + \hh y P_0 \delta  \theta ]$.
Keeping terms to linear order in the fluctuations  in~\eq{eq:dot-pa} 
and eliminating $\delta {P}$ in favor of  $\delta {\rho}$ 
yields
\begin{eqnarray}
&
\partial_t \delta \rho =
-  
 [v_1 \nabla_x \delta \rho +    v_0  \rho_0 \nabla_y \delta \theta], 
  \nonumber \\
&
\partial_t \delta \theta  =  - \lambda \rho_0  \nabla_x  \delta \theta   - 
\frac{v_0}{2\rho_0} \nabla_y \delta \rho   
  +     K_A   \nabla^2 \delta \theta  + \gamma     P^D_y  + \sqrt{\frac{ 2\Lambda }{ {\rho_0}}} f_y\;,
\label{eq:delta-theta}
\nonumber \\
&
\partial_t \rho^D =  -  v_D  \bm\nabla\cdot \delta \m P^D,
\\
& \partial_t  \delta \m P^D =   - D_{r} \delta  \m P^D 
 -  \frac{v_D}{2} \nabla  \delta \rho^D  + \sqrt{2\Lambda \rho^D_0}  \m f.
\nonumber
\end{eqnarray}
with {$v_1 = v_0\rho_0\alpha'(\rho_0)/[2 \alpha(\rho_0)]\simeq v_0 $, where the second equality holds deep in the flocking state.}
{We evaluate correlation functions in  Fourier space by introducing the Fourier amplitudes of the fluctuations,
 $g(\m q, \omega) = \int_{\m r,t} e^{-i (\omega t - \m q \cdot \m x)}  {g}(\m x, t) $, for any function $g$.
The angular correlations are then given by}
 \begin{equation}
\langle | {\theta} (\m q, \omega )|^2 \rangle = 
 \frac{  ( \omega + v_1  q_\parallel )^2
[ 
{2 \Lambda/\rho_0}
+ 
\gamma^2   \langle | {P}^D_y |^2 \rangle  
 ] 
  }{
( \omega + v_1  q_\parallel )^2   K^2_A   q^4 
+ 
 \left[
 ( \omega  +  v_1  q_\parallel ) 
(  \omega  +  \lambda \rho_0 q_\parallel )
   -     v^2_0  q^2_\perp/2 \right]^2
}
\label{eq:delta-theta2}
\end{equation}
where we let $\m q = q_\parallel \hh x + q_\perp \hh y$.
{The equations for the dissenters are decoupled form those of the aligners. The correlation function of the fluctuations in the dissenters' polarization density is then easily calculated, with the result }
{ 
 \begin{equation}
\fl  \langle  
   | {{P}}^D_x (\m q, \omega ) |^2
  \rangle
  =
\langle  
  | {{P}}^D_y (\m q, \omega ) |^2
  \rangle
= 
2 \rho^D_0 \Lambda 
\left[  \frac{ \hat{q}^2_\parallel   }{D^2_r +  \omega^2}
+  \frac{  \omega^2    \hat{q}^2_\perp
}{\omega^2 D^2_r + [ \omega^2 -\frac{v^2_D q ^2}{2} ]^2  }
\right],
\label{eq:P-Fourier}
  \end{equation}}
{ where $\hat{q}_{\parallel,\perp}=q_{\parallel,\perp}/q$.
The equal-time correlation is given by $\langle | {\theta} (q)|^2 \rangle=\int_\omega \langle | {\theta} (q, \omega )|^2 \rangle$, with the result}
 \begin{equation}
\langle |  {\theta}(\m q) |^2  \rangle = 
 \frac{  \Lambda}{\rho_0K_A   q^2_\parallel}  
  +\frac{ \Lambda \gamma^2 \rho^D_0
( D_{r}  +K_A  q^2_\parallel)  }
 { D_{r}K_A  q^2_\parallel  
 \left[(D_\mathrm{r} +K_A  q^2_\parallel )^2
+  \lambda^2\rho^2_0  q^2_\parallel \right]
} .
\label{eq:delta-theta23}
\end{equation}
{The first term on the right hand side of Eq.~(\ref{eq:delta-theta23}) are the fluctuations in the pure system, the second one is the contribution for the dissenters. In the long wavelength limit, and using that 
$\gamma =J a^2$, Eq. ~(\ref{eq:delta-theta23}) 
can be rewritten as \eq{eq:correlation-result}.
In other words the presence of dissenters essentially renormalizes the noise strength.
}

Finally, for comparison we note that static obstacles could be incorporated in the continuum model 
by quenched disorder corresponding to  a  stochastic force in \eq{eq:delta-theta2}  of the form
$\m F = -\beta_0 \bm{\nabla} \phi$ ,
 obtained as the gradient of 
a random potential $\phi$ (see, e.g.~\cite{Morin2017}) and 
 with correlations
$\langle F_i(\m x, t)  F_j(\m x', t')  \rangle = \beta^2_0  \nabla_i \nabla_j \delta(\m x-\m x') $.
Using this expression in  \eq{eq:delta-theta2} and considering the limit $k_\parallel \gg k_\perp$ as in~\cite{Morin2017}, 
 one obtains the result presented in the main text, ~\eq{eq:delta-theta3}.

\section*{References}
\providecommand{\newblock}{}


\begin{thebibliography}{10}
\expandafter\ifx\csname url\endcsname\relax
  \def\url#1{{\tt #1}}\fi
\expandafter\ifx\csname urlprefix\endcsname\relax\def\urlprefix{URL }\fi
\providecommand{\eprint}[2][]{\url{#2}}

\bibitem{Vicsek1995}
Vicsek T, Czir\'ok A, Ben-Jacob E, Cohen I and Shochet O 1995 {\em Phys. Rev.
  Lett.\/} {\bf 75}(6) 1226--1229

\bibitem{Bricard2013}
Bricard A, Caussin J~B, Desreumaux N, Dauchot O and Bartolo D 2013 {\em
  Nature\/} {\bf 503} 95--98

\bibitem{Copeland2009}
Copeland M and Weibel D~B 2009 {\em Soft Matter\/} {\bf 5} 1174--1187

\bibitem{Ballerini2008a}
Ballerini M, Cabibbo N, Candelier R, Cavagna A, Cisbani E, Giardina I, Lecomte
  V, Orlandi A, Parisi G, Procaccini A, Viale M and Zdravkovic V 2008 {\em
  Proc. Natl. Acad. Sci. USA\/} {\bf 105} 1232--1237

\bibitem{Silverberg2013}
Silverberg J~L, Bierbaum M, Sethna J~P and Cohen I 2013 {\em Phys. Rev.
  Lett.\/} {\bf 110}(22) 228701

\bibitem{Karamouzas2014}
Karamouzas I, Skinner B and Guy S~J 2014 {\em Phys. Rev. Lett.\/} {\bf 113}(23)
  238701 \urlprefix\url{http://link.aps.org/doi/10.1103/PhysRevLett.113.238701}

\bibitem{Chate2008}
Chat\'e H, Ginelli F, Gr\'egoire G and Raynaud F 2008 {\em Phys. Rev. E\/} {\bf
  77}(4) 046113

\bibitem{Solon2015c}
Solon A~P, Chat\'e H and Tailleur J 2015 {\em Phys. Rev. Lett.\/} {\bf 114}(6)
  068101

\bibitem{Solon2015d}
Solon A~P, Caussin J~B, Bartolo D, Chat\'e H and Tailleur J 2015 {\em Phys.
  Rev. E\/} {\bf 92}(6) 062111
  \urlprefix\url{http://link.aps.org/doi/10.1103/PhysRevE.92.062111}

\bibitem{Chepizhko2013}
Chepizhko O, Altmann E~G and Peruani F 2013 {\em Phys. Rev. Lett.\/} {\bf
  110}(23) 238101

\bibitem{Berdahl2013}
Berdahl A, Torney C~J, Ioannou C~C, Faria J~J and Couzin I~D 2013 {\em
  Science\/} {\bf 339} 574--576 ISSN 0036-8075
  \urlprefix\url{http://science.sciencemag.org/content/339/6119/574}

\bibitem{Morin2017}
Morin A, Desreumaux N, Caussin J~B and Bartolo D 2017 {\em Nat. Phys.\/} {\bf
  13} 63--67

\bibitem{Pince2016}
Pince E, Velu S~K~P, Callegari A, Elahi P, Gigan S, Volpe G and Volpe G 2016
  {\em Nat. Comm.\/} {\bf 7} 10907

\bibitem{Sandor2017}
S{\'a}ndor C, Lib{\'a}l A, Reichhardt C and Reichhardt C~J~O 2017 {\em Phys.
  Rev. E\/} {\bf 95} 032606

\bibitem{Guttal2010}
Guttal V and Couzin I~D 2010 {\em Proc. Natl. Acad. Sci. USA\/} {\bf 107}
  16172--16177
  \urlprefix\url{http://www.pnas.org/content/107/37/16172.abstract}

\bibitem{Couzin2011}
Couzin I~D, Ioannou C~C, Demirel G, Gross T, Torney C~J, Hartnett A, Conradt L,
  Levin S~A and Leonard N~E 2011 {\em Science\/} {\bf 334} 1578--1580 ISSN
  0036-8075 \urlprefix\url{http://science.sciencemag.org/content/334/6062/1578}

\bibitem{Baglietto2013}
Baglietto G, Albano E~V and Candia J 2013 {\em Physica A\/} {\bf 392}
  3240--3247

\bibitem{Ariel2015}
Ariel G, Rimer O and Ben-Jacob E 2015 {\em Journal of Statistical Physics\/}
  {\bf 158} 579--588 ISSN 1572-9613
  \urlprefix\url{http://dx.doi.org/10.1007/s10955-014-1095-7}

\bibitem{Copenhagen2016}
Copenhagen K, Quint D~A and Gopinathan A 2016 {\em Sci. Rep.\/} {\bf 6} 31808

\bibitem{Fily2012}
Fily Y and Marchetti M~C 2012 {\em Phys. Rev. Lett.\/} {\bf 108} 235702

\bibitem{Fily2014}
Fily Y, Henkes S and Marchetti M~C 2014 {\em Soft Matter\/} {\bf 10} 2132--40

\bibitem{Marchetti2016a}
Marchetti M, Fily Y, Henkes S, Patch A and Yllanes D 2016 {\em Curr. Opin.
  Colloid Interface Sci.\/} {\bf 21} 34--43

\bibitem{Szabo2006}
Szab\'o B, Sz\"oll\"osi G, G\"onci B, Jur\'anyi Z, Selmeczi D and Vicsek T 2006
  {\em Phys. Rev. E\/} {\bf 74} 061908

\bibitem{Henkes2011}
Henkes S, Fily Y and Marchetti M~C 2011 {\em Phys. Rev. E\/} {\bf 84} 040301

\bibitem{Weber2013}
Weber C~A, Hanke T, Deseigne J, L\'eonard S, Dauchot O, Frey E and Chat\'e H
  2013 {\em Phys. Rev. Lett.\/} {\bf 110}(20) 208001

\bibitem{Amit2005}
Amit D and Martin-Mayor V 2005 {\em Field Theory, the Renormalization Group,
  and Critical Phenomena\/} 3rd ed (Singapore: World Scientific)

\bibitem{Cooper1982}
Cooper F, Freedman B and Preston D 1982 {\em Nucl. Phys. B\/} {\bf 210} 210

\bibitem{Chen2012}
Chen X, Dong X, Be'er A, Swinney H~L and Zhang H~P 2012 {\em Phys. Rev.
  Lett.\/} {\bf 108}(14) 148101
  \urlprefix\url{http://link.aps.org/doi/10.1103/PhysRevLett.108.148101}

\bibitem{Peshkov2012}
Peshkov A, Ngo S, Bertin E, Chat\'e H and Ginelli F 2012 {\em Phys. Rev.
  Lett.\/} {\bf 109}(9) 098101
  \urlprefix\url{https://link.aps.org/doi/10.1103/PhysRevLett.109.098101}

\bibitem{Attanasi2014}
Attanasi A, Cavagna A, Del~Castello L, Giardina I, Melillo S, Parisi L, Pohl O,
  Rossaro B, Shen E, Silvestri E and Viale M 2014 {\em Phys. Rev. Lett.\/} {\bf
  113}(23) 238102

\bibitem{Baglietto2008}
Baglietto G and Albano E~V 2008 {\em Phys. Rev. E\/} {\bf 78}(2) 021125

\bibitem{Farrell2012}
Farrell F~D~C, Marchetti M~C, Marenduzzo D and Tailleur J 2012 {\em Phys. Rev.
  Lett.\/} {\bf 108} 1--5

\bibitem{Toner2012}
Toner J, Tu Y and Ramaswamy S 2005 {\em Annals of Physics\/} {\bf 318} 170--244

\bibitem{Couzin2005}
Couzin I~A, Krause J, Franks N~R and Levin S~A 2005 {\em Nature\/} {\bf 433}
  513--516

\bibitem{Quint2015}
Quint D and Gopinathan A 2015 {\em Physical Biology\/} {\bf 17} 046008

\bibitem{Bertin09}
Bertin E, Droz M and Gr\'egoire G 2009 {\em J. Phys. A: Math. Theor.\/} {\bf
  42} 445001

\end{thebibliography}
\end{document}